\documentclass[11pt]{article}

\usepackage{amsmath,amsfonts,amssymb,amsthm}
\usepackage{mathrsfs}
\usepackage{color}

\definecolor{labelkey}{gray}{.8}
\definecolor{refkey}{gray}{.8}

\definecolor{darkred}{rgb}{0.9,0.1,0.1}
\definecolor{darkgreen}{rgb}{0,0.5,0}

\newtheorem{theorem}{Theorem}[section]
\newtheorem{lemma}[theorem]{Lemma}

\newtheorem{proposition}[theorem]{Proposition}

\theoremstyle{remark}

\numberwithin{equation}{section}
\newcommand{\R}{\mathbb{R}}

\newcommand{\E}{\mathbb{E}}
\newcommand{\F}{\mathcal{F}}

\newcommand{\G}{\mathcal{G}}

\newcommand{\V}{\mathcal{V}}

\newcommand{\RR}{\mathcal{R}}
\newcommand{\eps}{\varepsilon}

\newcommand{\g}{\mathfrak{g}}

\newcommand{\farc}{\frac}

\begin{document}

\title{The random Schr\"odinger equation: slowly decorrelating
  time-dependent potentials}

\author{Yu Gu\thanks{Department of Mathematics, Stanford University, Stanford, CA, 94305, USA. 
Email: yg@math.stanford.edu; ryzhik@math.stanford.edu.}  \and Lenya Ryzhik\footnotemark[1]}
\date{}
\maketitle

\begin{abstract}
We analyze the weak-coupling limit of the random Schr\"odinger equation with low frequency initial data and a slowly 
decorrelating random potential. For the probing signal with a sufficiently long wavelength, we prove a homogenization result, that is, 
the properly compensated wave field admits a deterministic limit in the ``very low" frequency regime. The limit is
``anomalous" in the sense that the solution behaves as $\exp(-Dt^{s})$ with $s>1$ rather than 
the ``usual"~$\exp(-Dt)$ homogenized behavior when the random potential is rapidly decorrelating.
Unlike in rapidly decorrelating potentials, as we decrease the wavelength of the probing signal, 
stochasticity appears in the asymptotic limit -- there exists a critical scale depending on the random potential which separates 
the deterministic and stochastic regimes.
\end{abstract}

\section{Introduction}

We consider the weakly random Schr\"odinger equation 
\begin{equation}\label{june1602}
i\partial_t\phi+\frac12\Delta\phi-\eps V(t,x)\phi=0
\end{equation}
with a low frequency initial condition $\phi(0,x)=\phi_0(\ell x)$. Here, $\eps\ll 1$ is a small parameter measuring the strength of the random
potential $V(t,x)$, and $\ell\ll 1$ is the ratio of the typical scale of variations of the potential to that of the initial data. We are interested in
the long time behavior of the solution, on the scales such that the effect of the weak random potential is visible. 
When the temporal and spatial
correlations of the random potential are decaying rapidly, this problem was addressed in \cite{bkr} when $\ell=1$ (so that the initial condition
is not slowly varying), and in~\cite{bckr,gr2015} for~$\ell\ll 1$. We should also mention the papers~\cite{Erdos-Yau2,Spohn} 
where the kinetic limit for the case $\ell=1$ is obtained in the much harder case of \emph{time-independent} random potentials. 
In the aforementioned papers, with the rapid decay of the 
correlations of the random potential, the solution of (\ref{june1602}) 
is affected by the random potential in a non-trivial way on a universal time scale $t\sim\eps^{-2}$ which  depends neither  on 
the typical scale of variations of the initial data nor on the covariance structure of the random potential. Moreover, the limit is deterministic
for all $\ell\ll 1$ -- all solutions with slowly varying initial data homogenize at times $t\sim\eps^{-2}$.
Here, we are interested in what happens when the correlations of the random potential decay slowly. 
As we will see, then the time scale to observe ``non-trivial" behaviors indeed depends on the correlations of the 
random potential and is not universal. Moreover, the observed behavior on this time scale varies dramatically depending on 
the scale of variations of the initial condition even for slowly varying initial data.

Let us now be more specific about our assumptions on the random potential $V(t,x)$.
It is a stationary mean-zero Gaussian random field, with the spectral representation
\begin{equation}
V(t,x)=\int_{\R^d}e^{ip\cdot x}\frac{\tilde{V}(t,dp)}{(2\pi)^d} 
\end{equation}
and the covariance function
\begin{equation}\label{june1606}
R(t,x)=\E\{V(t+s,x+y)V(s,y)\}= \int_{\R^d}\hat{R}(p)e^{-\g(p) t}e^{ip\cdot x}\frac{dp}{(2\pi)^d}.
\end{equation}
Here,
\begin{equation}
\hat{R}(p)=\frac{a(p)}{|p|^{2\gamma+d-2}}, ~~~\g(p)=\mu|p|^{2\beta},
\end{equation}
are, respectively, the spatial power spectrum of  the potential and its spectral gap.
The cut-off function $a(p)\geq 0$ is bounded and compactly supported, and $\mu>0$ is a constant. 
The role of the parameters $\gamma>0$ and $\beta>0$ can be seen by setting $t=0$ and $x=0$, respectively, in
(\ref{june1606}):
\[
R(0,x)=\int_{\R^d}e^{ip\cdot x}\frac{a(p)}{|p|^{2\gamma+d-2}}\frac{dp}{(2\pi)^d}
\sim \frac{1}{|x|^{2-2\gamma}},~~|x|\gg 1,
\]
and
\[
R(t,0)=\int_{\R^d}\ e^{-\mu|p|^{2\beta} t}\frac{a(p)}{|p|^{2\gamma+d-2}} 
\frac{dp}{(2\pi)^d}\sim\frac{1}{|t|^{(2-2\gamma)/(2\beta)}},~~~|t|\gg 1.
\]
One of the main results of~\cite{gr2015} is that in the case $\gamma+\beta<1$ (rapidly decorrelating potentials),  
on the time scale $t\sim\eps^{-2}$, and
after a proper phase compensation, the wave field  is homogenized -- it is nearly deterministic, as long as $\ell\ll 1$. 
In this paper, we consider the opposite regime: the  parameters~$\gamma$ and $\beta$ satisfy
\begin{equation}\label{june1604}
0<\gamma,\beta<1,~~\gamma+\beta>1,
\end{equation} 
so that the correlation function decays slowly.
Our goal is to explore the asymptotic behavior of the wave field, and its
dependence on the the initial condition.

We mention that the case $\ell=1$, that is, when the  the probing signal
varies on a scale comparable to that of the random potential, was investigated in \cite{bkr}. It was shown that, 
after  the phase compensation, the wave field behaves on an anomalous time scale like a lognormal distribution. 

\subsubsection*{The possible asymptotic limits}

Before presenting our main results, let us give a heuristic explanation of what one may expect. From now on we assume
that $\ell=\eps^\alpha$, and investigate different values of $\alpha>0$.
Consider a time scale $\eps^{-\kappa}$ with some $\kappa>0$ to be determined.
The rescaled wave function 
\[
\phi_\eps(t,x)=\phi(\farc{t}{\eps^\kappa},\frac{x}{\eps^\alpha}),
\]
satisfies the rescaled Schr\"odinger equation 
\begin{equation}
i\partial_t \phi_\eps(t,x)+\frac12\eps^{2\alpha-\kappa}\Delta \phi_\eps-
\frac{1}{\eps^{\kappa-1}}V(\frac{t}{\eps^\kappa},\frac{x}{\eps^\alpha})\phi_\eps(t,x)=0,~~~\phi_\eps(0,x)=\phi_0(x).
\label{eq:reEq}
\end{equation}
The modified potential 
\begin{equation}
V_\eps(t,x)=\frac{1}{\eps^{\kappa-1}}V(\frac{t}{\eps^\kappa},\frac{x}{\eps^\alpha})
\end{equation} 
has the covariance function
\begin{equation}\label{july3002}
R_\eps(t,x)=\frac{1}{\eps^{2\kappa-2}}\int_{\R^d}\frac{a(p)}{|p|^{2\gamma+d-2}}e^{-\mu|p|^{2\beta}t/\eps^\kappa}
e^{ip\cdot x/\eps^\alpha}\frac{dp}{(2\pi)^d}
=\int_{\R^d}\frac{a(p\eps^{\kappa/2\beta})}{|p|^{2\gamma+d-2}}e^{-\mu|p|^{2\beta}t} e^{i(p\cdot x)  \eps^{\frac{\kappa}{2\beta}-\alpha}}\frac{dp}{(2\pi)^d},
\end{equation}
which is of order $O(1)$ provided that the time scale exponent is chosen as
\begin{equation}
\kappa=\frac{2\beta}{2\beta+\gamma-1}.
\end{equation}
It follows from (\ref{june1604}) that
$1<\kappa<2$, hence the time scale $\eps^{-\kappa}$ is much shorter than the classical central limit theorem time
scale $\eps^{-2}$. We also see that for $\alpha=\alpha_c$, with
\begin{equation}
\alpha_c=\frac{\kappa}{2\beta}=\frac{1}{2\beta+\gamma-1},
\end{equation}
the rescaled potential $V_\eps(t,x)$ has the same distribution as $V(t,x)$ modulo the effects from the cut-off~$a(p)$.
The self-similar structure of the power spectrum   plays an important role in our analysis.  

Let us ignore, for the moment, the effects of the term $\eps^{2\alpha-\kappa}\Delta$ in \eqref{eq:reEq}, so that this equation 
formally reduces to an ODE.  
If  $\alpha=\alpha_c$, the rescaled covariance function has a limit,
\begin{equation}
R_\eps(t,x)\to \bar R(t,x)=\int_{\R^d}\frac{a(0)}{|p|^{2\gamma+d-2}}e^{-\mu|p|^{2\beta}t}e^{ip\cdot x}\frac{dp}{(2\pi)^d}
\end{equation}
as $\eps\to 0$,
which can be identified as the covariance function of a generalized Gaussian random field~$\dot{W}(t,x)$. 
That is, there exists a generalized fractional Gaussian random field $\dot{W}(t,x)$ so that
\begin{equation}\label{aug302}
\E\{\dot{W}(t+s,x+y)\dot{W}(s,y)\}=\frac{1}{(2\pi)^d}\int_{\R^d}\frac{a(0)}{|p|^{2\gamma+d-2}}e^{-\mu|p|^{2\beta}t}e^{ip\cdot x}dp.
\end{equation}
Thus, when $\alpha=\alpha_c$, for each $x\in\R^d$ we formally have
\begin{equation}
\partial_t \phi_\eps(t,x)\approx-i\dot{W}(t,x)\phi_\eps(t,x)
\end{equation}
when $\eps$ is small, so the solution is approximately
\begin{equation}
\phi_\eps(t,x)\approx  \phi_0(x)\exp\Big\{-i\int_0^t \dot{W}(s,x)ds\Big\}.
\end{equation}
Furthermore, when $\alpha\in (0,\alpha_c)$, the limit of the covariance function is
\begin{equation}
R_\eps(t,x)\to \frac{1}{(2\pi)^d}\int_{\R^d}\frac{a(0)}{|p|^{2\gamma+d-2}}e^{-\mu|p|^{2\beta}t}dp,
\end{equation}
so the spatial variable is ``frozen" and there is only temporal mixing. Then, \eqref{eq:reEq} becomes, formally:
\begin{equation}
\partial_t \phi_\eps(t,x)\approx-i\dot{W}(t,0)\phi_\eps(t,x)
\end{equation}
with the approximate solution 
\begin{equation}
\phi_\eps(t,x)\approx \phi_0(x)\exp\Big\{-i\int_0^t \dot{W}(s,0)ds\Big\}.
\end{equation}
Its Fourier transform 
\[
\hat{\phi}_0(\xi) \exp\Big\{-i\int_0^t \dot{W}(s,0)ds\Big\}
\]
 coincides with the limit in the case $\alpha=0$ \cite[Theorem 1.2]{bkr}. 

On the other hand, when $\alpha>\alpha_c$, the covariance (\ref{july3002}) is highly oscillatory, and so is the corresponding
noise $\dot{W}(t,x/\eps^{\alpha-\alpha_c})$, and it is not clear on this
formal level what the limit should be. It turns out that the oscillations will lead to a deterministic limit.

\subsection*{The main result}

We analyze the problem in the Fourier domain, where \eqref{eq:reEq} can be written as
\[
i\partial_t \hat{\phi}_\eps-\frac12\eps^{2\alpha-\kappa}|\xi|^2\hat{\phi}_\eps-\widehat{V_\eps \phi_\eps}=0.
\]
The initial profile
$\phi_0$ is assumed to be of the Schwartz class:~$\phi_0\in \mathcal{S}(\R^d)$.
As in~\cite{bkr,gr2015}, in order to eliminate the large phase coming simply from the deterministic evolution,
we consider the compensated wave function\begin{equation}
\psi_\eps(t,\xi)=\hat{\phi}_\eps(t,\xi)e^{\frac{i|\eps^\alpha\xi|^2t}{2\eps^\kappa}}=\eps^{\alpha d}\hat{\phi}(\frac{t}{\eps^\kappa},\eps^\alpha\xi)e^{\frac{i|\eps^\alpha\xi|^2t}{2\eps^\kappa}}.
\label{eq:repsi}
\end{equation}
Let us emphasize that $\xi\sim O(1)$ in the argument of the function $\psi_\eps(t,\xi)$ corresponds 
to $\xi\sim O(\eps^\alpha)$ in the argument of the function $\hat{\phi}(t,\xi)$. In the following, 
if we refer to the order of frequencies, it is with respect to the argument of $\hat{\phi}$.

In order to formulate the main result, we define the following constants
\begin{equation*}
K_1=\Omega_d\int_0^\infty e^{-\mu \rho^{2\beta}}\frac{d\rho}{\rho^{2\gamma-1}},
\end{equation*}
where $\Omega_d$ is the surface area of the unit sphere in $\R^d$, and
\begin{eqnarray*}
&&K_2(\lambda,\xi)=\int_{\R^d}e^{-\mu|w|^{2\beta}}
\frac{e^{iw\cdot \xi \lambda^{1-(2\beta)^{-1}}}}{|w|^{2\gamma+d-2}}\frac{dw}{(2\pi)^d},\\
&&D=\frac{a(0)K_1\kappa^2}{(2\pi)^d(2-\kappa)},
\\
&&
D(t,\xi)=\frac{a(0)}{(2\pi)^d}\int_{[0,1]^2}|s-u|^{-\frac{1-\gamma}{\beta}}K_2(|s-u|t,\xi)dsdu.
\end{eqnarray*}
We denote by $N(0,\sigma^2)$ a random variable with the normal distribution of mean zero and variance~$\sigma^2$, and by 
$\dot W(t,x)$ a generalized fractional Gaussian mean-zero random field with covariance (\ref{aug302}).
\begin{theorem}
For each fixed $t>0,\xi\in \R^d$, the compensated wave function $\psi_\eps(t,\xi)$ converges in distribution, as $\eps\to 0$,
to a limit $\bar{\psi}(t,\xi)$ defined as follows: (i) if $\alpha>\alpha_c$, then the limit is deterministic:
\[
\bar{\psi}(t,\xi)= \hat{\phi}_0(\xi)\exp\Big\{-\frac12Dt^{{2}/{\kappa}}\Big\},
\]
(ii) for $\alpha=\alpha_c$ the limit is random:
\[
\bar{\psi}(t,\xi)= \int_{\R^d}\phi_0(x)e^{-i\xi\cdot x}\exp\Big\{-i\int_0^t \dot{W}(s,x)ds\Big\} dx,
\]
(iii) if $0<\alpha<\alpha_c$ and $\beta\in(0,1/2]$, or $\kappa-\alpha_c<\alpha<\alpha_c$ and $\beta\in(1/2,1)$, then
the limit is random:
\[
\bar{\psi}(t,\xi)=\hat{\phi}_0(\xi)\exp\big\{iN(0,D t^{{2}/{\kappa}})\big\},
\]
(iv) if $\alpha=\kappa-\alpha_c$ and $\beta\in(1/2,1)$, then 
\[
\bar{\psi}(t,\xi)=\hat{\phi}_0(\xi)\exp\big\{iN(0,D(t,\xi)t^{{2}/{\kappa}})\big\}.
\]
\label{thm:mainTH}
\end{theorem}
We point out that when $\alpha\in (\alpha_c,\infty)$, the limit in Theorem~\ref{thm:mainTH} 
is deterministic, thus we have a convergence in probability. This means
that solutions with ``sufficiently low frequency" initial data homogenize also in the random potentials with slowly decaying 
correlations. The (non-standard) time factor $t^{2/\kappa}$ in the exponent   reflects the slowly decorrelating
nature of the random media, and should be contrasted with the exponential decay in time $\exp(-Dt)$ 
of the homogenized limit
in rapidly decorrelating potentials~\cite{gr2015}. It is also worth noting that the wave field is attenuated in the limit and the exponential decay in time implies a loss of mass (the $L^2$ norm). The same phenomenon occurs in the rapidly decorrelating 
potentials~\cite{gr2015}, where the   mass lost from the low frequencies of the order $O(\eps^\alpha)$ escapes to 
the high frequencies of the order $O(1)$.  

A key feature of Theorem~\ref{thm:mainTH} is the existence of a critical wavelength scale $O(\eps^{-\alpha_c})$
of the probing signal, which  separates the deterministic and random regimes. This is very different from the Schr\"odinger equation with
rapidly decorrelating potentials, where homogenization happens for all slowly varying initial data (any $\alpha>0$),
 as was proved in~\cite{gr2015}. 
In addition, the random variable~$N(0,Dt^{2/\kappa})$ that arises in the limit for $\alpha<\alpha_c$ can be identified as the (one-point) 
distribution of a fractional Brownian motion at 
time $t$, which may be written as 
\[
\sqrt{D}B_{1/\kappa}(t)=\int_0^t \dot{W}(s,0)ds,
\]
in agreement with our informal analysis. Such fractional limits are typical for additive functionals of 
Gaussian random variables with slowly decaying  correlations, but it is not a universal limit 
and comes from the specific covariance structure of the Gaussian field~\cite{taqqu1975weak}. A posteriori, this limit justifies the basic assumption of the informal computation we have shown above:
the effects of the Laplacian operator are suppressed via the phase compensation, the dynamics is essentially reduced to an ODE, 
and the random potential behaves as a fractional Gaussian noise in the limit. 
We should mention that the restriction~$\alpha\geq \kappa-\alpha_c$ for $\beta\in(1/2,1)$
is a limitation of the technique of the proof, and is the analog of the restriction $\beta\le 1/2$ that was needed in~\cite{bkr}
in the case $\alpha=0$ considered there. 


A heuristic explanation of homogenization in the ``very low frequency" regime ($\alpha>\alpha_c$) may be as follows. 
Up to the time scale $\eps^{-\kappa}$, the characteristic frequencies of the slowly decorrelating medium are of 
the order $O(\eps^{\alpha_c})$ -- this can be seen from expression (\ref{july3002}) for the rescaled covariance~$R_\eps(t,x)$. 
For the probing signal with a bandwidth narrower than $O(\eps^{\alpha_c})$, 
of the order $O(\eps^\alpha)$ with~$\alpha>\alpha_c$, the interactions of the signal with the medium frequencies can only produce
frequencies outside of the original bandwidth, so only the ballistic component survives in the limit, leading to the homogenization result.
On the other hand, if the probing signal signal has bandwidth wider than~$O(\eps^{\alpha_c})$, that is, $\alpha<\alpha_c$, 
then the interactions with the medium frequencies
can produce frequencies inside the initial bandwidth, and the limit is stochastic.

To the best of our knowledge, the first study of wave propagation in slowly decorrelating media was done in 
the one-dimensional case \cite{garnier2009pulse,marty2009acoustic}, where it was shown that a pulse going through a random medium
with long-range correlation performs a fractional Brownian motion around its mean position, as opposed to the regular 
Brownian motion in the rapidly decorrelating case \cite{FGPS-07}. On the other hand, the 
motion of particles in such random media leading to fractional Brownian limits was considered in \cite{fannjiang2000fractional,komorowski2007asymptotics,komorowski2007passive}. 
We also mention the recent 
work of \cite{gomez2012radiative,gomez2013wave} analyzing the wave energy instead of phase evolution, 
where a time-separation is observed due to the long-range correlations in the random potential.

The paper is organized as follows. In Section~\ref{s:dex}, we introduce the Duhamel expansion and prove some basic moment estimates. 
Next, we prove the homogenization result in Section~\ref{s:hom}. The discussion of the stochastic regimes is in Section~\ref{s:sto}.

{\bf Acknowledgment.} This work was supported by the AFOSR NSSEFF Fellowship and
NSF grant DMS-1311903.

\section{The Duhamel expansion and a moment estimate}
\label{s:dex}

The basic strategy in the passage to the limit $\eps\to 0$ is the same as for rapidly decorrelating potentials in~\cite{gr2015}.
We consider the Duhamel expansion for the function $\psi_\eps$ and study its moments using the expansion. In this section,
we introduce the series and establish a uniform (in~$\eps\in(0,1)$) bound on the moments of the terms in the expansion. This will allow 
us to pass to the limit term-wise
in the series for any moment of  the compensated wave function.

\subsubsection*{The Duhamel expansion}

A straightforward calculation shows that the compensated wave function
$\psi_\eps$ defined in~\eqref{eq:repsi} satisfies the following integral equation:
\begin{equation}
\psi_\eps(t,\xi)=\hat{\phi}_0(\xi)+\frac{\eps}{i\eps^\kappa}\int_0^t
\int_{\R^d}\frac{\tilde{V}(\frac{s}{\eps^\kappa},dp)}{(2\pi)^d}e^{i(|\eps^\alpha \xi|^2-|\eps^\alpha\xi-p|^2)s/2\eps^\kappa} \psi_\eps(s,\xi-\frac{p}{\eps^\alpha})ds.
\label{eq:ineq}
\end{equation}
The solution of~\eqref{eq:ineq} can be written 
as an iterated series:  
\begin{equation}\label{july3104}
\psi_\eps(t,\xi)=\sum_{n=0}^\infty f_{n,\eps}(t,\xi),
\end{equation}
with the individual terms
\begin{equation}\label{july3102}
f_{n,\eps}(t,\xi)=\left(\frac{\eps}{i\eps^\kappa}\right)^n \int_{\Delta_n(t)}\int_{\R^{nd}}\prod_{j=1}^n \frac{\tilde{V}(\frac{s_j}{\eps^\kappa},dp_j)}{(2\pi)^d} e^{i G_n(\eps^\alpha \xi, s^{(n)},p^{(n)})/\eps^\kappa} \hat{\phi}_0(\xi-\frac{p_1+\ldots+p_n}{\eps^\alpha}).
\end{equation}
The phase factor in (\ref{july3102}) is
\begin{equation}
G_n(\xi, s^{(n)},p^{(n)})=
\sum_{k=1}^n (|\xi-p_1-\ldots-p_{k-1}|^2-|\xi-p_1-\ldots-p_k|^2)\frac{s_k}{2}.
\label{eq:phase}
\end{equation}
We used here the convention $f_{0,\eps}(t,\xi)=\hat{\phi}_0(\xi)$,
and have set $p_0=0$, $p^{(n)}=(p_1,\ldots,p_n)$,
as well as~$s^{(n)}=(s_1,\ldots,s_n)$.
We have also defined the time simplex
\[
\Delta_n(t)=\{0\leq s_n\leq\ldots\leq s_1\leq t\}.
\]

As has been shown in, for instance,~\cite{bkr,gr2015}, the series (\ref{july3104})
converges, and one can take the expectation of $\psi_\eps$ and its moments
term-wise, as long as $\eps>0$ is fixed. 
Thus, the  proof of Theorem~\ref{thm:mainTH} boils down to the 
asymptotic analysis of the moments of the form 
\begin{equation}\label{july3106}
\E\{ f_{m_1,\eps}\ldots f_{m_M,\eps}  f_{n_1,\eps}^*\ldots f_{n_N,\eps}^*\}.
\end{equation}
As $V$ is a mean-zero Gaussian field, 
a non-zero contribution comes only from the terms with
\begin{equation}\label{july3110}
\sum_{i=1}^M m_i+\sum_{j=1}^N n_j=2k
\end{equation}
for some $k\in \mathbb{N}$. 

We will denote ``the random part" in (\ref{july3106}) as 
\begin{equation*}
\begin{aligned}
I_{M,N}=&(2\pi)^{-2kd}\tilde{V}(\frac{s_{1,1}}{\eps^\kappa},dp_{1,1})\ldots 
\tilde{V}(\frac{s_{1,m_1}}{\eps^\kappa},dp_{1,m_1})\ldots 
\tilde{V}(\frac{s_{M,1}}{\eps^\kappa},dp_{M,1})\ldots
\tilde{V}(\frac{s_{M,m_M}}{\eps^\kappa},dp_{M,m_M})\\
&\times\tilde{V}^*(\frac{u_{1,1}}{\eps^\kappa},dq_{1,1})\ldots 
\tilde{V}^*(\frac{u_{1,n_1}}{\eps^\kappa},dq_{1,n_1})\ldots 
\tilde{V}^*(\frac{u_{N,1}}{\eps^\kappa},dq_{N,1})
\ldots\tilde{V}^*(\frac{u_{N,n_N}}{\eps^\kappa},dq_{N,n_N}).
\end{aligned}
\end{equation*}
Here, $s_j$ and $p_j$ variables come from the terms $f_{m_k,\eps}$, -- each of them 
involves $m_k$ temporal variables~$s_{k,1},\dots, s_{k,m_k}$ and $m_k$ momentum variables
$p_{k,1},\dots, p_{k,m_k}$, while the variables $u_k$ and 
$q_k$ come from the terms $f_{n_k,\eps}^*$ in (\ref{july3106}).
Using the rules of computing the $2k-$th joint moment of 
mean zero Gaussian random variables, we write
\begin{equation}
\E\{I_{M,N}\}=\sum_{\F} \prod_{(v_l,v_r)\in\F} 
e^{-\g(w_l)|v_{l}-v_{r}|/\eps^\kappa}\delta(w_l+ w_r)\hat{R}(w_l)
\farc{dw_ldw_r}{(2\pi)^{d}} .
\label{eq:IMN}
\end{equation}
The summation $\sum_\F$ extends over all pairings ${\cal F}$ formed
over 
the vertices 
\[
\{s_{1,1},\ldots,s_{1,m_1},\ldots,s_{M,1},\ldots,s_{M,m_M}u_{1,1},
\ldots,u_{1,n_1},\ldots,u_{N,1},\ldots,u_{N,n_N}\}.
\]
In (\ref{eq:IMN}), $v_l,v_r$ are the two vertices of a given pair, and
$w_l,w_r$ are the corresponding $p,q$ variables, that is, $w_l=p_{i,j}$ if
$v_l=s_{i,j}$ and $w_l=-q_{i,j}$ if $v_l=u_{i,j}$. The same holds for
$w_r$. We will also write a pair as an edge $e=(v_l,v_r)$. Note that the
order of $v_l,v_r$ does not matter here since both $\g$ and~$\hat{R}$ are 
even. 

\subsubsection*{A uniform bound on the individual terms} 

As we have mentioned, 
in order to be able to pass to the limit term-wise in the series for the moments of $\psi_\eps$ we will need the following uniform
bound.
\begin{lemma}\label{lem:bdmm}
For all $\eps\in (0,1]$, we have
\begin{equation}
|\E\{ f_{m_1,\eps}\ldots f_{m_M,\eps}  f_{n_1,\eps}^*\ldots f_{n_N,\eps}^*\}|\leq \frac{(2k-1)!!}{\prod_{i=1}^M (m_i)!\prod_{j=1}^N (n_j)!}C^k
\end{equation}
with some constant $C$ depending on $t,\phi_0,\hat{R},\g$, and $k$ as in (\ref{july3110}). 
\end{lemma}

{\bf Proof.} 
Since $\hat{\phi}_0$ is bounded, we have
\begin{equation*}
|\E\{ f_{m_1,\eps}\ldots f_{m_M,\eps}  f_{n_1,\eps}^*\ldots f_{n_N,\eps}^*\}|\leq  C^{k}\eps^{(1-\kappa)2k}
 \int_{\Delta_{m,n}(t)}dsdu \int_{\R^{2kd}} |\E\{I_{M,N}\}|.
\end{equation*}
where
\[
\Delta_{m,n}(t)=\Delta_{m_1}(t)\times\ldots\times \Delta_{n_N}(t).
\]
By symmetry, the r.h.s. of the above expression is bounded by
\[
\begin{aligned}
&\frac{C^{k}\eps^{2(1-\kappa)k}}{\prod_{i=1}^M(m_i)!\prod_{j=1}^N (n_j)!} \int_{[0,t]^{2k}}dsdu \int_{\R^{2kd}}|\E\{I_{M,N}\}|\\
=&\frac{C^{k}(2k-1)!!}{\prod_{i=1}^M(m_i)!\prod_{j=1}^N(n_j)!}\left(\eps^{2(1-\kappa)} \int_{[0,t]^2}\int_{\R^d}\frac{e^{-\g(w)|s-u|/\eps^\kappa}}{(2\pi)^d}\hat{R}(w)dwdsdu\right)^k,
\end{aligned}
\]
with the factor $(2k-1)!!$ coming from the total number of pairings. We recall that $\g(p)=\mu|p|^{2\beta}$ and 
\[
\hat{R}(p)=\frac{a(p)}{|p|^{2\gamma+d-2}},
\] 
and change the variable
\[
w\mapsto \frac{\eps^{\kappa/2\beta}}{|s-u|^{1/2\beta}}w
\]
 to obtain
\begin{equation*}
\eps^{2(1-\kappa)} \int_{[0,t]^2}\int_{\R^d}\frac{e^{-\g(w)|s-u|/\eps^\kappa}}{(2\pi)^d}\hat{R}(w)dwdsdu
= \int_{[0,t]^2}\int_{\R^d}\frac{e^{-\mu |w|^{2\beta}}}{(2\pi)^d|w|^{2\gamma+d-2}} 
a(\frac{\eps^{\kappa/2\beta}w}{|s-u|^{1/2\beta}})\frac{dwdsdu}{|s-u|^{\frac{1-\gamma}{\beta}}} ,
\end{equation*}
which is bounded since $a$ is bounded, $\gamma<1$ and $\gamma+\beta>1$,
and the conclusion of Lemma~\ref{lem:bdmm} follows.~$\Box$
 
The bound in Lemma~\ref{lem:bdmm} is useful, since for fixed $M,N$, we have -- recall, once again, that $k$ is related to 
$m_i$ and $n_i$ via (\ref{july3110}):
\begin{equation}
\sum_{m_1=0}^\infty \ldots \sum_{n_N=0}^\infty 
\frac{(2k-1)!!}{\prod_{i=1}^M (m_i)!
\prod_{j=1}^N (n_j)!}C^k=\sum_{k=0}^\infty 
\frac{(2k-1)!!C^k(M+N)^{2k}}{(2k)!}<\infty,
\end{equation}
as can be seen by the binomial expansion. Thus, Lemma~\ref{lem:bdmm} implies 
that the individual terms in the series for 
$|\E\{ \psi_\eps(t,\xi)^M\psi_\eps^*(t,\xi)^N\}|$ are uniformly bounded in 
$\eps\in (0,1]$ by a summable series, and thus not only we may write
\begin{equation}
\begin{aligned}
\E\{ \psi_\eps(t,\xi)^M\psi_\eps^*(t,\xi)^N\}=\sum_{m_1=0}^\infty \ldots \sum_{n_N=0}^\infty\E\{ f_{m_1,\eps}\ldots f_{m_M,\eps}  f_{n_1,\eps}^*\ldots f_{n_N,\eps}^*\}
\end{aligned}
\end{equation}
but we may also pass to the limit individually in each summand.

\subsubsection*{Re-writing the individual terms}

We will now look at the individual
term 
\[
\E\{ f_{m_1,\eps}\ldots f_{m_M,\eps}  f_{n_1,\eps}^*\ldots f_{n_N,\eps}^*\},
\]
and re-write in a form convenient for the analysis. 
We can write
\begin{equation}
\begin{aligned}
\E\{ f_{m_1,\eps}\ldots f_{m_M,\eps}  f_{n_1,\eps}^*\ldots f_{n_N,\eps}^*\}=&
\left(\frac{\eps}{i\eps^\kappa}\right)^{\sum_{i=1}^Mm_i}\left(\frac{\eps}{-i\eps^\kappa}\right)^{\sum_{j=1}^N n_j}\\
\times & \int_{\Delta_{m,n}(t)}dsdu\int_{\R^{2kd}}\E\{I_{M,N}\}e^{i\G_M}e^{-i\G_N} \prod_{i=1}^M h_{M,i}\prod_{j=1}^N h_{N,j}^*,
\end{aligned}
\end{equation}
where  
\[
\G_M=\sum_{i=1}^M G_{m_i}(\eps^\alpha \xi,s^{(m_i)},p^{(m_i)})/\eps^\kappa, 
~~~\G_N=\sum_{j=1}^N G_{n_j}(\eps^\alpha \xi, u^{(n_j)},q^{(n_j)})/\eps^\kappa,
\]
\[
h_{M,i}=\hat{\phi}_0(\xi-\frac{p_{i,1}+\ldots+p_{i,m_i}}{\eps^\alpha}), h_{N,j}^*=
\hat{\phi}_0^*(\xi-\frac{q_{j,1}+\ldots+q_{j,n_j}}{\eps^\alpha}),
\]
and we recall that
\[
\Delta_{m,n}(t)=\Delta_{m_1}(t)\times\ldots\times \Delta_{n_N}(t).
\]
By \eqref{eq:IMN}, we   have 
\begin{equation}
\E\{I_{M,N}\}=\sum_{\F} \prod_{(v_l,v_r)\in\F}  
e^{-\g(w_l)|v_{l}-v_{r}|/\eps^\kappa}\delta(w_l+ w_r)\hat{R}(w_l)
\farc{dw_ldw_r}{(2\pi)^{d}},
\end{equation}
where $\F$ are the pairings arising in computing the joint moments of the Gaussians $\tilde V$.
Thus, we can write
\begin{equation}
\E\{ f_{m_1,\eps}\ldots f_{m_M,\eps}  f_{n_1,\eps}^*\ldots f_{n_N,\eps}^*\}=
\sum_\F J^\eps_{m_1,\ldots,n_N^*}(\F),
\end{equation}
with the individual terms
\begin{equation}
\begin{aligned}
&J^\eps_{m_1,\ldots,n_N^*}(\F)=\left(\frac{\eps}{i\eps^\kappa}\right)^{\sum_{i=1}^Mm_i}
\left(\frac{\eps}{-i\eps^\kappa}\right)^{\sum_{j=1}^N n_j}\\
\times & \int_{\Delta_{m,n}(t)}dsdu\int_{\R^{2kd}}
\prod_{(v_l,v_r)\in\F}  
e^{-\g(w_l)|v_{l}-v_{r}|/\eps^\kappa}\delta(w_l+ w_r)\hat{R}(w_l)
e^{i\G_M}e^{-i\G_N} \prod_{i=1}^M h_{M,i}\prod_{j=1}^N h_{N,j}^*
\frac{dw_ldw_r}{(2\pi)^{d}}.
\label{eq:defJ}
\end{aligned}
\end{equation}
Our goal is, therefore to identify 
\[
\lim_{\eps\to0} J^\eps_{m_1,\ldots,n_N^*}(\F),
\]
for a fixed pairing $\F$.
We integrate out the $w_r$ variables 
in \eqref{eq:defJ} and obtain
\begin{equation}
\begin{aligned}
&J^\eps_{m_1,\ldots,n_N^*}(\F)=
\left(\frac{\eps}{i\eps^\kappa}\right)^{\sum_{i=1}^Mm_i}
\left(\frac{\eps}{-i\eps^\kappa}\right)^{\sum_{j=1}^N n_j}\\
\times & \int_{\Delta_{m,n}(t)}dsdu
\int_{\R^{kd}}\prod_{(v_l,v_r)\in\F}  
e^{-\g(w_l)|v_{l}-v_{r}|/\eps^\kappa}\hat{R}(w_l) 
e^{i\G_M}e^{-i\G_N} \prod_{i=1}^M h_{M,i}\prod_{j=1}^N h_{N,j}^*
\frac{dw_l}{(2\pi)^{d}},
\end{aligned}
\end{equation}
with $\G_M,\G_N,h_{M,i},h_{N,j}$ subjected to the
constraint 
\begin{equation}\label{july3112}
w_l+w_r=0
\end{equation}
for all pairs.
Once again, we recall that 
\[
\g(p)=\mu|p|^{2\beta},~~\hat{R}(p)=\farc{a(p)}{|p|^{2\gamma+d-2}}.
\]
Using these expressions, for every $w_l$, we apply a key rescaling
to extract the characteristic frequencies from the underlying 
media up to a time scale $\eps^{-\kappa}$:
\[
w_l\mapsto  \eps^{\alpha_c} w_l.
\]
Using the fact that $\alpha_c=\kappa/2\beta$ and 
$\kappa=2\beta/(2\beta+\gamma-1)$, we obtain
\begin{equation}
\begin{aligned}
&J^\eps_{m_1,\ldots,n_N^*}(\F)=\frac{1}{i^{\sum_{i=1}^Mm_i}}
\frac{1}{(-i)^{\sum_{j=1}^N n_j}}\\
\times & \int_{\Delta_{m,n}(t)}dsdu\int_{\R^{kd}}
\prod_{(v_l,v_r)\in\F}  
e^{-\mu|w_l|^{2\beta}|v_{l}-v_{r}|}
\frac{a(\eps^{\alpha_c}w_l)}{|w_l|^{2\gamma+d-2}} 
e^{i\tilde{\G}_M}e^{-i\tilde{\G}_N} \prod_{i=1}^M \tilde{h}_{M,i}\prod_{j=1}^N \tilde{h}_{N,j}^*\frac{dw_l}{(2\pi)^{d}},
\end{aligned}
\label{eq:defJc}
\end{equation}
where
\begin{equation}
\tilde{\G}_M=\sum_{i=1}^M G_{m_i}(\eps^\alpha \xi,s^{(m_i)},\eps^{\alpha_c}p^{(m_i)})/\eps^\kappa, \G_N=\sum_{j=1}^N G_{n_j}(\eps^\alpha \xi, u^{(n_j)},\eps^{\alpha_c}q^{(n_j)})/\eps^\kappa,
\label{eq:phasec}
\end{equation}
and
\begin{equation}
\tilde{h}_{M,i}=\hat{\phi}_0(\xi-\eps^{\alpha_c-\alpha}(p_{i,1}+\ldots+p_{i,m_i})), \tilde{h}_{N,j}^*=\hat{\phi}_0^*(\xi-\eps^{\alpha_c-\alpha}(q_{j,1}+\ldots+q_{j,n_j})),
\label{eq:icc}
\end{equation}
which are still subjected to the constraints (\ref{july3112}) for all edges.
This expression will be the starting point for our analysis for all frequencies. 


\section{Homogenization of the very low frequencies}
\label{s:hom}

We begin with the very low frequency regime ($\alpha>\alpha_c$),
where the limit does not depend on whether~$\beta\le 1/2$ or $\beta>1/2$,
and prove the homogenization result -- the limit of the compensated 
wave function is deterministic.  
\begin{proposition}
If $\alpha>\alpha_c$, then, for a fixed $t>0$ and $\xi\in\R^d$, we have
\[
\psi_\eps(t,\xi)\to \hat{\phi}_0(\xi)e^{-\frac12D t^{{2}/{\kappa}}}
\] 
in probability as $\eps\to 0$.
\label{prop:homo}
\end{proposition}
{\bf Proof.}
We will first compute the limit of $\E\{\psi_\eps(t,\xi)\}$, and then show that the second absolute moment converges to the square of the limit of the first 
moment, which will establish the convergence in probability.

\subsubsection*{The limit of the first moment}

Let us start with $\E\{\psi_\eps(t,\xi)\}$. In that case, there is only one simplex of time variables, and
\eqref{eq:defJc} simplifies to
\[
J_n^\eps(\F)=\frac{1}{i^n}\hat{\phi}_0(\xi)\int_{\Delta_{2k}(t)}ds 
\int_{\R^{kd}}\prod_{(v_l,v_r)\in\F}  
e^{-\mu|w_l|^{2\beta}|v_{l}-v_{r}|}
\frac{a(\eps^{\alpha_c}w_l)}{|w_l|^{2\gamma+d-2}}
e^{iG_n(\eps^\alpha \xi,s^{(n)},\eps^{\alpha_c}p^{(n)})/\eps^\kappa}
\frac{dw_l}{(2\pi)^{d}},
\]
with $n=2k$ for some $k\in \mathbb{N}$. By the definition \eqref{eq:phase} of $G_n$, and since  
$\alpha>\alpha_c=\kappa/2\beta$, we have a bound for the phase factor 
\[
|G_n(\eps^\alpha \xi,s^{(n)},\eps^{\alpha_c}p^{(n)})/\eps^\kappa| \leq C \eps^{\frac{\kappa}{\beta}-\kappa},
\]
with a constant $C$ depending on $\xi,s^{(n)},p^{(n)}$. Since $\beta<1$, the function $a(p)$ is uniformly bounded and the function
\[
e^{-\mu|p|^{2\beta}|s-u|}/|p|^{2\gamma+d-2},
\]
is in $L^1([0,t]^2\times \R^d)$,
we may apply the Lebesgue dominated convergence theorem. This gives 
\[
J_{2k}^\eps(\F)\to\frac{1}{i^{2k}}\hat{\phi}_0(\xi)\int_{\Delta_{2k}(t)}ds \int_{\R^{kd}}\prod_{(v_l,v_r)\in\F}
 e^{-\mu|w_l|^{2\beta}|v_{l}-v_{r}|}\frac{a(0)}{|w_l|^{2\gamma+d-2}}\farc{dw_l}{(2\pi)^{d}}.
\]
Therefore, we have
\[
\begin{aligned}
\lim_{\eps\to 0}\E\{\psi_\eps(t,\xi)\}=&\sum_{k=0}^\infty  \sum_\F \lim_{\eps\to 0}J_{2k}^\eps (\F)\\
=& \sum_{k=0}^\infty \sum_\F(-1)^k\hat{\phi}_0(\xi)\int_{\Delta_{2k}(t)}ds \int_{\R^{kd}}
\prod_{(v_l,v_r)\in\F}  e^{-\mu|w_l|^{2\beta}|v_{l}-v_{r}|}\frac{a(0)}{|w_l|^{2\gamma+d-2}}\farc{dw_l}{(2\pi)^{d}}.
\end{aligned}
\]
Integrating out the $w_l$ variables, and recalling the definition of
\[
K_1=\Omega_d\int_0^\infty e^{-\mu \rho^{2\beta}}\frac{d\rho}{\rho^{2\gamma-1}},
\]  we obtain
\[
\lim_{\eps\to 0}\E\{\psi_\eps(t,\xi)\}= \hat{\phi}_0(\xi) \sum_{k=0}^\infty(-1)^k
\Big[ \farc{a(0)K_1}{(2\pi)^{d}}\Big]^k 
\sum_\F\int_{\Delta_{2k}(t)}ds  \prod_{(v_l,v_r)\in\F}|v_l-v_r|^{-\frac{1-\gamma}{\beta}}.
\]
Note that 
\[
\sum_\F \prod_{(v_l,v_r)\in\F} |v_l-v_r|^{-\frac{1-\gamma}{\beta}}
\] 
is symmetric in $(s_1,\ldots,s_{2k})$, so
\[
\sum_\F \int_{\Delta_{2k}(t)}ds\prod_{(v_l,v_r)\in\F}|v_l-v_r|^{-\frac{1-\gamma}{\beta}}=\frac{(2k-1)!!}{(2k)!}\left(\int_{[0,t]^2}|s-u|^{-\frac{1-\gamma}{\beta}}dsdu\right)^k,
\]
and thus
\[
\lim_{\eps\to 0}\E\{\psi_\eps(t,\xi)\}= \hat{\phi}_0(\xi)e^{-\frac12D t^{{2}/{\kappa}}}.
\]

\subsubsection*{The limit of the second absolute moment}

Next, we consider $\E\{|\psi_\eps(t,\xi)|^2\}$, then \eqref{eq:defJc} becomes
\[
\begin{aligned}
J^\eps_{m,n^*}(\F)=&\frac{1}{i^m}\frac{1}{(-i)^n}\int_{\Delta_{m,n}(t)}dsdu\int_{\R^{kd}}
\prod_{(v_l,v_r)\in\F}  e^{-\mu|w_l|^{2\beta}|v_{l}-v_{r}|}\frac{a(\eps^{\alpha_c}w_l)}{|w_l|^{2\gamma+d-2}}\farc{dw_l}{(2\pi)^{d}} \\
\times& e^{iG_m(\eps^\alpha \xi, s^{(m)}, \eps^{\alpha_c}p^{(m)})/\eps^\kappa}e^{-iG_n(\eps^\alpha \xi, u^{(n)}, \eps^{\alpha_c}q^{(n)})/\eps^\kappa} \hat{\phi}_0(\xi-\eps^{\alpha_c-\alpha}\sum_{i=1}^m p_i)\hat{\phi}_0^*(\xi-\eps^{\alpha_c-\alpha}\sum_{j=1}^n q_j),
\end{aligned}
\]
with $m+n=2k$ for some $k\in\mathbb{N}$. Here, the $p,q$ variables satisfy the constraint 
\[
w_l+w_r=0
\]
for all edges. 

The ``crossing" pairings   that  have at least one edge $e=(v_l,v_r)$ such that $v_l$ is an $s$-variable, and $v_r$ is a~$u$-variable satisfy
\[
\sum_{i=1}^m p_i=-\sum_{j=1}^n q_j\neq 0.
\]
Thus, such pairings give the following factor coming from the initial condition 
\[
 \hat{\phi}_0(\xi-\eps^{\alpha_c-\alpha}\sum_{i=1}^m p_i)\hat{\phi}_0(\xi-\eps^{\alpha_c-\alpha}\sum_{j=1}^n q_j)\to 0
 \]
 as $\eps\to 0$ , as $\alpha>\alpha_c$. If we denote the set of those ``crossing" pairings by $P_c$, then it is clear that
 \[
 \lim_{\eps\to 0} \sum_{\F\in P_c}J_{m,n^*}^\eps(\F)=0.
 \]

For $\F\notin P_c$, all edges $e=(v_l,v_r)$ connect the same type of variables, thus both~$m$ and~$n$ are even, 
and 
\[
\sum_{i=1}^m p_i=\sum_{j=1}^n q_j=0,
\]
under the constraint $w_l+w_r=0$. Then, by the same argument as for $\E\{\psi_\eps(t,\xi)\}$, we have
 \[
 \begin{aligned}
 J_{m,n^*}^\eps(\F)\to&\frac{1}{i^m}\frac{1}{(-i)^n}|\hat{\phi}_0|^2(\xi)\int_{\Delta_{m,n}(t)}dsdu\int_{\R^{kd}}
 \prod_{(v_l,v_r)\in\F}  e^{-\mu|w_l|^{2\beta}|v_{l}-v_{r}|}\frac{a(0)}{|w_l|^{2\gamma+d-2}}\farc{dw_l}{(2\pi)^{d}} \\
 =&\frac{1}{i^m}\frac{1}{(-i)^n}|\hat{\phi}_0|^2(\xi)\Big[\frac{a(0)K_1}{(2\pi)^{d}}\Big]^k
 \int_{\Delta_{m,n}(t)}dsdu\prod_{(v_l,v_r)\in\F}|v_l-v_r|^{-\frac{1-\gamma}{\beta}}.
 \end{aligned}
 \]
 Since $m,n$ are both even, we write $m=2k_1,n=2k_2$ for $k_1,k_2\in\mathbb{N}$, and obtain
 \[
 \begin{aligned}
 &\lim_{\eps\to 0}\E\{|\psi_\eps(t,\xi)|^2\}=\sum_{k_1,k_2=0}^\infty\ \sum_{\F\notin P_c} \lim_{\eps\to 0}J_{2k_1,2k_2^*}^\eps(\F)\\
 =&\sum_{k_1,k_2=0}^\infty(-1)^{k_1+k_2}|\hat{\phi}_0|^2(\xi)\Big[\frac{a(0)K_1}{(2\pi)^{d}}\Big]^{k_1+k_2} 
  \sum_{\F\notin P_c}\int_{\Delta_{m,n}(t)}dsdu\prod_{(v_l,v_r)\in\F}|v_l-v_r|^{-\frac{1-\gamma}{\beta}}.
 \end{aligned}
 \]
As $\F\notin P_c$, we have 
 \[
 \begin{aligned}
 & \sum_{\F\notin P_c}\int_{\Delta_{m,n}(t)}dsdu\prod_{(v_l,v_r)\in\F}|v_l-v_r|^{-\frac{1-\gamma}{\beta}}\\
 =&\left(\sum_{\F_s}\int_{\Delta_{2k_1}(t)}ds \prod_{(v_l,v_r)\in \F_s}|v_l-v_r|^{-\frac{1-\gamma}{\beta}}\right)\times \left(\sum_{\F_u}\int_{\Delta_{2k_2}(t)}du \prod_{(v_l,v_r)\in \F_u}|v_l-v_r|^{-\frac{1-\gamma}{\beta}}\right),
 \end{aligned}
 \]
 where $\F_s,\F_u$ denote the pairings formed by $s_1,\ldots,s_{2k_1}$ and $u_1,\ldots,u_{2k_2}$ respectively. This implies
\[
\lim_{\eps\to 0}\E\{|\psi_\eps(t,\xi)|^2\}
=|\lim_{\eps\to 0} \E\{\psi_\eps(t,\xi)\}|^2,
\]
which completes the proof.~$\Box$


\section{The stochastic limits}
\label{s:sto}

In this section, we assume $\alpha\in (0,\alpha_c]$ and prove a convergence in law of the compensated
wave function $\psi_\eps(t,\xi)$.
To identify the limiting distribution, we compute the limiting moments. 
Recall that by Lemma~\ref{lem:bdmm} we only need to pass to the limit in \eqref{eq:defJc}:
\begin{equation}
\begin{aligned}
&J^\eps_{m_1,\ldots,n_N^*}(\F)=\frac{1}{i^{\sum_{i=1}^Mm_i}}\frac{1}{(-i)^{\sum_{j=1}^N n_j}}\\
\times & \int_{\Delta_{m,n}(t)}dsdu\int_{\R^{kd}}\prod_{(v_l,v_r)\in\F}  
e^{-\mu|w_l|^{2\beta}|v_{l}-v_{r}|}\frac{a(\eps^{\alpha_c}w_l)}{|w_l|^{2\gamma+d-2}}\farc{dw_l}{(2\pi)^{d}}
 e^{i\tilde{\G}_M}e^{-i\tilde{\G}_N} \prod_{i=1}^M \tilde{h}_{M,i}\prod_{j=1}^N \tilde{h}_{N,j}^*.
\end{aligned}
\label{eq:defJc1}
\end{equation}
The phase factors are
\[
\tilde{\G}_M=\sum_{i=1}^M G_{m_i}(\eps^\alpha \xi,s^{(m_i)},\eps^{\alpha_c}p^{(m_i)})/\eps^\kappa, ~~~~~
\tilde{\G}_N=\sum_{j=1}^N G_{n_j}(\eps^\alpha \xi, u^{(n_j)},\eps^{\alpha_c}q^{(n_j)})/\eps^\kappa,
\]
thus for $\alpha\leq\alpha_c$, we have
\[
|\tilde{\G}_M|+|\tilde{\G}_N|\sim \eps^{\alpha+\alpha_c-\kappa}.
\]
We will need to consider two cases. 

\textbf{Case 1: $\beta\in (0,1/2]$}. Then $\alpha_c\ge \kappa$, thus we have 
\[
\alpha+\alpha_c-\kappa>0,
\]
which implies $\tilde{\G}_M,\tilde{\G}_N\to 0$ as $\eps\to 0$, so by \eqref{eq:defJc1}
\begin{equation}
\begin{aligned}
J^\eps_{m_1,\ldots,n_N^*}(\F)\to& \frac{1}{i^{\sum_{i=1}^Mm_i}}\frac{1}{(-i)^{\sum_{j=1}^N n_j}}\\
&\times  \int_{\Delta_{m,n}(t)}dsdu\int_{\R^{kd}}\prod_{(v_l,v_r)\in\F} 
e^{-\mu|w_l|^{2\beta}|v_{l}-v_{r}|}\frac{a(0)}{|w_l|^{2\gamma+d-2}}\frac{dw_l}{(2\pi)^{d}} H_\alpha(\xi,p,q),
\end{aligned}
\label{eq:limitJ1}
\end{equation}
with 
\begin{eqnarray}\label{aug402}
&&H_\alpha(\xi,p,q)=\lim_{\eps\to 0}\prod_{i=1}^M \tilde{h}_{M,i}\prod_{j=1}^N \tilde{h}_{N,j}^*\\
&&~~~~~~~~~~~~~~=\left\{\begin{array}{ll}
(\hat{\phi}_0(\xi))^M(\hat{\phi}_0^*(\xi))^N & \alpha\in (0,\alpha_c),\\
\prod_{i=1}^M \hat{\phi}_0(\xi-p_{i,1}-\ldots-p_{i,m_i})\prod_{j=1}^N \hat{\phi}_0^*(\xi-q_{j,1}-\ldots-q_{j,n_j}) &\alpha=\alpha_c.
\end{array}
\right.\nonumber
\end{eqnarray}
We recall that all $p,q-$variables satisfy the constraints $w_l+w_r=0$ for all edges.

\textbf{Case 2: $\beta\in(1/2,1)$}. Here, we will consider the range 
\[
\kappa-\alpha_c<\alpha\le\alpha_c.
\]
We have 
 \[
 \alpha+\alpha_c-\kappa> 0 \mbox{ for } 
 \alpha\in (\kappa-\alpha_c,\alpha_c],
 \]
or
 \[
\alpha+\alpha_c-\kappa=0 \mbox{ for } \alpha=\kappa-\alpha_c.
 \]
When $\alpha\in (\kappa-\alpha_c,\alpha_c]$, we still have
\[
\hbox{$\tilde{\G}_M,\tilde{\G}_N\to 0$ as $\eps\to 0$},
\]
so we have the same limit as \eqref{eq:limitJ1}:
 \begin{equation}
\begin{aligned}
J^\eps_{m_1,\ldots,n_N^*}(\F)\to& \frac{1}{i^{\sum_{i=1}^Mm_i}}\frac{1}{(-i)^{\sum_{j=1}^N n_j}}\\
&\times  \int_{\Delta_{m,n}(t)}dsdu\int_{\R^{kd}}\prod_{(v_l,v_r)\in\F}  
e^{-\mu|w_l|^{2\beta}|v_{l}-v_{r}|}\frac{a(0)}{|w_l|^{2\gamma+d-2}}\farc{dw_l}{(2\pi)^{d}} H_\alpha(\xi,p,q).
\end{aligned}
\label{eq:limitJ2}
\end{equation} 
On the other hand, when $\alpha=\kappa-\alpha_c$,  we have
 \[
 \tilde{\G}_M\to \sum_{i=1}^M \sum_{k=1}^{m_i} (p_{i,k}\cdot \xi) s_{i,k} \mbox{ and }  
 \tilde{\G}_N\to \sum_{j=1}^N \sum_{k=1}^{n_j} (q_{j,k}\cdot \xi) u_{j,k},
 \]
hence
\begin{equation}
\begin{aligned}
J^\eps_{m_1,\ldots,n_N^*}(\F)\to& \frac{1}{i^{\sum_{i=1}^Mm_i}}\frac{1}{(-i)^{\sum_{j=1}^N n_j}}
\int_{\Delta_{m,n}(t)}dsdu\int_{\R^{kd}}\prod_{(v_l,v_r)\in\F}  
e^{-\mu|w_l|^{2\beta}|v_{l}-v_{r}|}\frac{a(0)}{|w_l|^{2\gamma+d-2}}\farc{dw_l}{(2\pi)^{d}} \\
&\times e^{i\sum_{i=1}^M \sum_{k=1}^{m_i} (p_{i,k}\cdot \xi) s_{i,k}} 
e^{-i  \sum_{j=1}^N \sum_{k=1}^{n_j} (q_{j,k}\cdot \xi) u_{j,k}} (\hat{\phi}_0(\xi))^M(\hat{\phi}_0^*(\xi))^N.
\label{eq:limitJ3}
\end{aligned}
\end{equation}
To summarize, depending on the parameters $\alpha,\beta,\gamma$ ($\kappa$ depends on $\beta,\gamma$), we can write
 \[
 \lim_{\eps\to 0}J_{m_1,\ldots,n_N^*}^\eps(\F)=J_{m_1,\ldots,n_N^*}(\F),
 \]
 with $J_{m_1,\ldots,n_N^*}(\F)$ given by \eqref{eq:limitJ1}, \eqref{eq:limitJ2} and \eqref{eq:limitJ3} in the three respective cases:
 (i) $0<\beta\le 1/2$ and~$0<\alpha\le\alpha_c$, (ii) $1/2<\beta<1$ and $\kappa-\alpha_c<\alpha\le\alpha_c$; and (iii)
 $1/2<\beta<1$ and $\alpha=\kappa-\alpha_c$.

 \subsubsection*{ The case $\alpha<\alpha_c$}
 
Theorem~\ref{thm:mainTH} treats two regimes in the case $0<\alpha<\alpha_c$:
(i) $\beta\in (0,1/2]$ and $\alpha\in (0,\alpha_c)$, and (ii)~$\beta\in (1/2,1)$, and $\alpha\in [\kappa-\alpha_c,\alpha_c)$.
The key feature in this case is that the factor (\ref{aug402}) coming from the initial conditions in \eqref{eq:defJc1} has the limit
\[
\prod_{i=1}^M \tilde{h}_{M,i}\prod_{j=1}^N \tilde{h}_{N,j}^*\to (\hat{\phi}_0(\xi))^M(\hat{\phi}_0^*(\xi))^N
\]
 as $\eps\to 0$.
If, in addition, we assume that either $\beta\in(0,1/2]$ (case (i) above),
or $\beta\in(1/2,1)$ but~$\alpha\neq\kappa-\alpha_c$ in case (ii), then
the term $J$ also simplifies: the phase information vanishes in the limit, and
\begin{eqnarray}\label{eq:limitJnp}
&&J_{m_1,\ldots,n_N^*}(\F)= \frac{1}{i^{\sum_{i=1}^Mm_i}}\frac{1}{(-i)^{\sum_{j=1}^N n_j}}\\
&&~~~~~~~~~~~~~~~~~
\times  \int_{\Delta_{m,n}(t)}\!dsdu\int_{\R^{kd}}\!\prod_{(v_l,v_r)\in\F}  
e^{-\mu|w_l|^{2\beta}|v_{l}-v_{r}|}\frac{a(0)}{|w_l|^{2\gamma+d-2}}\frac{dw_l}{(2\pi)^{d}}
 (\hat{\phi}_0(\xi))^M(\hat{\phi}_0^*(\xi))^N.
\nonumber 
\end{eqnarray}
Making a change of variable $w_l\mapsto w_l|v_l-v_r|^{-1/2\beta}$ and integrating out $w_l$, we obtain
\[
\begin{aligned}
J_{m_1,\ldots,n_N^*}(\F)=& \frac{1}{i^{\sum_{i=1}^Mm_i}}\frac{1}{(-i)^{\sum_{j=1}^N n_j}}\\
&\times\Big [\frac{a(0)K_1}{(2\pi)^{d}}\Big]^k(\hat{\phi}_0(\xi))^M(\hat{\phi}_0^*(\xi))^N
\int_{\Delta_{m,n}(t)}dsdu \prod_{(v_l,v_r)\in\F}|v_l-v_r|^{-\frac{1-\gamma}{\beta}}.
\end{aligned}
\]
As in the  proof of Proposition~\ref{prop:homo}, we use symmetry after summing over all pairings $\F$ to get
\[
\begin{aligned}
\sum_\F J_{m_1,\ldots,n_N^*}(\F)=&\frac{1}{i^{\sum_{i=1}^Mm_i}}
\frac{1}{(-i)^{\sum_{j=1}^N n_j}}\Big [\frac{a(0)K_1}{(2\pi)^{d}}\Big]^k(\hat{\phi}_0(\xi))^M(\hat{\phi}_0^*(\xi))^N\\
&\times  \frac{(2k-1)!!}{\prod_{i=1}^M m_i!\prod_{j=1}^N n_j!}\left(\int_{[0,t]^2}|s-u|^{-\frac{1-\gamma}{\beta}}dsdu\right)^k,
\end{aligned}
\]
which implies
\[
\begin{aligned}
\lim_{\eps\to 0}\E\{\psi_\eps(t,\xi)^M\psi_\eps^*(t,\xi)^N\}=\sum_{m_1=0}^\infty \ldots \sum_{n_N=0}^\infty 
&\frac{1}{i^{\sum_{i=1}^Mm_i}}\frac{1}{(-i)^{\sum_{j=1}^N n_j}}
\Big [\frac{a(0)K_1}{(2\pi)^{d}}\Big]^k(\hat{\phi}_0(\xi))^M(\hat{\phi}_0^*(\xi))^N\\
\times & \frac{(2k-1)!!}{\prod_{i=1}^M m_i!\prod_{j=1}^N n_j!}\left(\int_{[0,t]^2}|s-u|^{-\frac{1-\gamma}{\beta}}dsdu\right)^k,
\end{aligned}
\]
with 
\[
2k=\sum_{i=1}^M m_i+\sum_{j=1}^N n_j
\]
in the summand. The binomial expansion tells us that
\[
\sum_{m_1+\ldots+n_N=2k} \frac{1}{i^{\sum_{i=1}^Mm_i}}\frac{1}{(-i)^{\sum_{j=1}^N n_j}}\frac{(2k)!}{\prod_{i=1}^M m_i!\prod_{j=1}^N n_j!}=(\frac{M}{i}-\frac{N}{i})^{2k}=(M-N)^{2k}(-1)^k,
\]
thus we have
\begin{eqnarray*}
&&\lim_{\eps\to 0}\E\{\psi_\eps(t,\xi)^M\psi_\eps^*(t,\xi)^N\}=\sum_{k=0}^\infty (M-N)^{2k}(-1)^k
\Big [\frac{a(0)K_1}{(2\pi)^{d}}\Big]^k
(\hat{\phi}_0(\xi))^M(\hat{\phi}_0^*(\xi))^N\\
&&\times \frac{1}{2^kk!}\left(\int_{[0,t]^2}|s-u|^{-\frac{1-\gamma}{\beta}}dsdu\right)^k
=(\hat{\phi}_0(\xi))^M(\hat{\phi}_0^*(\xi))^N\exp\Big\{-\frac12(M-N)^2D t^{{2}/{\kappa}}\Big\}\\
&&=\E\Big\{\left(\hat{\phi}_0(\xi) e^{iN(0, Dt^{{2}/{\kappa}})}\right)^M\left(\hat{\phi}_0^*(\xi) e^{-iN(0, Dt^{{2}/{\kappa}})}\right)^N
\Big\},
\end{eqnarray*}
which completes the proof in the cases $\beta\in(0,1/2], \alpha\in (0,\alpha_c)$ and $\beta\in(1/2,1),\alpha\in(\kappa-\alpha_c,\alpha_c)$. 

In the case $\beta\in(1/2,1), \alpha=\kappa-\alpha_c$, the phase does not disappear in the limit, and we have
\begin{equation}
\begin{aligned}
J_{m_1,\ldots,n_N^*}(\F)=& \frac{1}{i^{\sum_{i=1}^Mm_i}}\frac{1}{(-i)^{\sum_{j=1}^N n_j}}
\int_{\Delta_{m,n}(t)}dsdu\int_{\R^{kd}}\prod_{(v_l,v_r)\in\F} 
e^{-\mu|w_l|^{2\beta}|v_{l}-v_{r}|}\frac{a(0)}{|w_l|^{2\gamma+d-2}}\frac{dw_l}{(2\pi)^{d}}  \\
&\times e^{i\sum_{i=1}^M \sum_{k=1}^{m_i} (p_{i,k}\cdot \xi) s_{i,k}} 
e^{-i  \sum_{j=1}^N \sum_{k=1}^{n_j} (q_{j,k}\cdot \xi) u_{j,k}} (\hat{\phi}_0(\xi))^M(\hat{\phi}_0^*(\xi))^N.
\label{eq:limitJp}
\end{aligned}
\end{equation}
Compared to \eqref{eq:limitJnp}, the only difference is the extra phase factor 
\[
\sum_{i=1}^M \sum_{k=1}^{m_i} (p_{i,k}\cdot \xi) s_{i,k}-\sum_{j=1}^N \sum_{k=1}^{n_j} (q_{j,k}\cdot \xi )u_{j,k}.
\]
Since the $p,q-$variables in the above expression satisfy the constraint 
\[
w_l+w_r=0,
\]
we can write \eqref{eq:limitJp} as
\begin{equation*}
\begin{aligned}
J_{m_1,\ldots,n_N^*}(\F)=& \frac{1}{i^{\sum_{i=1}^Mm_i}}\frac{1}{(-i)^{\sum_{j=1}^N n_j}}
\int_{\Delta_{m,n}(t)}dsdu\int_{\R^{kd}}\prod_{(v_l,v_r)\in\F}  e^{-\mu|w_l|^{2\beta}|v_{l}-v_{r}|}
\frac{a(0)e^{i(w_l\cdot \xi)|v_l-v_r|}}{|w_l|^{2\gamma+d-2}}\frac{dw_l}{(2\pi)^{d}} \\
&\times (\hat{\phi}_0(\xi))^M(\hat{\phi}_0^*(\xi))^N.
\end{aligned}
\end{equation*}
Once again, we may change the variable $w_l\mapsto w_l|v_l-v_r|^{-1/2\beta}$ and integrate out $w_l$ to obtain
\[
\begin{aligned}
J_{m_1,\ldots,n_N^*}(\F)=& \frac{1}{i^{\sum_{i=1}^Mm_i}}\frac{1}{(-i)^{\sum_{j=1}^N n_j}}
[a(0)]^k(\hat{\phi}_0(\xi))^M(\hat{\phi}_0^*(\xi))^N\\
\times &
\int_{\Delta_{m,n}(t)}dsdu\prod_{(v_l,v_r)\in\F}|v_l-v_r|^{-\frac{1-\gamma}{\beta}}K_2(|v_l-v_r|,\xi).
\end{aligned}
\]
We recall that
\begin{equation}
K_2(\lambda,\xi)=\int_{\R^d}e^{-\mu |w|^{2\beta}}
\frac{e^{i(w\cdot \xi) \lambda^{1-(2\beta)^{-1}}}}{|w|^{2\gamma+d-2}}
\frac{dw}{(2\pi)^d}.
\end{equation}
By symmetry we have, after summing over all pairings
\[
\begin{aligned}
\sum_\F J_{m_1,\ldots,n_N^*}(\sigma,\F)=&\frac{1}{i^{\sum_{i=1}^Mm_i}}\frac{1}{(-i)^{\sum_{j=1}^N n_j}}
[a(0)]^k(\hat{\phi}_0(\xi))^M(\hat{\phi}_0^*(\xi))^N\\
&\times \frac{(2k-1)!!}{\prod_{i=1}^M m_i!\prod_{j=1}^N n_j!}\left(\int_{[0,t]^2}|s-u|^{-\frac{1-\gamma}{\beta}}
K_2(|s-u|,\xi)dsdu\right)^k,
\end{aligned}
\]
and as before, this implies
\begin{eqnarray*}
&&\lim_{\eps\to 0}\E\{\psi_\eps(t,\xi)^M\psi_\eps^*(t,\xi)^N\}=\sum_{k=0}^\infty (M-N)^{2k}(-1)^k
[a(0)]^k(\hat{\phi}_0(\xi))^M(\hat{\phi}_0^*(\xi))^N\\
&&\times \frac{1}{2^kk!}\left(\int_{[0,t]^2}|s-u|^{-\frac{1-\gamma}{\beta}}K_2(|s-u|,\xi)dsdu\right)^k
=(\hat{\phi}_0(\xi))^M(\hat{\phi}_0^*(\xi))^Ne^{-\frac12(M-N)^2D(t,\xi)t^{{2}/{\kappa}}}.
\end{eqnarray*}
The proof in the case $\beta\in(1/2,1),\alpha=\kappa-\alpha_c$ is now complete.

\subsubsection*{The case $\alpha=\alpha_c$}
 
When $\alpha=\alpha_c$, the argument in the initial condition in the expression for $J$ is different:
\begin{equation}
\begin{aligned}
J_{m_1,\ldots,n_N^*}(\F)=&\frac{1}{i^{\sum_{i=1}^Mm_i}}\frac{1}{(-i)^{\sum_{j=1}^N n_j}}  
\int_{\Delta_{m,n}(t)}dsdu\int_{\R^{kd}}\prod_{(v_l,v_r)\in\F}  
e^{-\mu|w_l|^{2\beta}|v_{l}-v_{r}|}\frac{a(0)}{|w_l|^{2\gamma+d-2}}\frac{dw_l}{(2\pi)^{d}}\\
&\times \prod_{i=1}^M \hat{\phi}_0(\xi-p_{i,1}-\ldots-p_{i,m_i})\prod_{j=1}^N \hat{\phi}_0^*(\xi-q_{j,1}-\ldots-q_{j,n_j}).
\label{eq:limitJpc}
\end{aligned}
\end{equation}
The goal is to show that 
\[
\lim_{\eps\to 0}\E\{\psi_\eps(t,\xi)^M\psi_\eps^*(t,\xi)^N\}=\sum_{m_1=0}^\infty\ldots\sum_{n_N=0}^\infty  
\sum_\F J_{m_1,\ldots,n_N^*}(\F)=\E\{\bar{\psi}(t,\xi)^M\bar{\psi}^*(t,\xi)^N\},
\]
where
\[
\bar{\psi}(t,\xi)=\int_{\R^d}\phi_0(x)e^{-i\xi\cdot x}\exp\Big\{-i\int_0^t \dot{W}(s,x)ds\Big\} dx,
\]
and $\dot{W}(s,x)$ is a generalized Gaussian random field with the covariance
\[
\E\{\dot{W}(s,x)\dot{W}(u,y)\}=\RR(s-u,x-y)= 
\int_{\R^d} e^{-\mu|w|^{2\beta}|s-u|}\frac{a(0)}{|w|^{2\gamma+d-2}}e^{iw\cdot (x-y)}\frac{dw}{(2\pi)^d}.
\]
Let us define
\begin{equation}
\bar{f}_n(t,\xi)=\frac{1}{n!}\int_{\R^d}\phi_0(x)(-i\V_t(x))^n e^{-i\xi\cdot x}dx
\end{equation}
with 
\begin{equation}
\V_t(x)=\int_0^t\dot{W}(s,x)ds.
\end{equation} 
The following lemma will help us express the moments of $\bar{\psi}$ as a series expansion.
\begin{lemma}
We have 
\[
\bar{\psi}(t,\xi)=\sum_{n=0}^\infty \bar{f}_n(t,\xi),
\]
and 
\begin{equation}
\E\{\bar{\psi}(t,\xi)^M\bar{\psi}^*(t,\xi)^N\}=\sum_{m_1=0}^\infty\ldots \sum_{n_N=0}^\infty \E\{ \bar{f}_{m_1}\ldots \bar{f}_{m_M}\bar{f}_{n_1}^*\ldots\bar{f}_{n_N}^*\}.
\label{eq:mmre}
\end{equation}
\end{lemma}
{\bf Proof.}
For every $t>0$ fixed,  $\V_t(x)$ is a mean-zero stationary spatial Gaussian random field, so for fixed $t>0$ and $x\in\R^d$, we have
\[
\sum_{n=0}^N \frac{1}{n!} (-i\V_t(x))^n\to e^{-i\V_t(x)}
\]
in $L^p(\Omega)$ as $N\to\infty$ with the $L^p$ error independent of $x$ for any $p\geq 1$. This implies 
\[
\sum_{n=0}^N \bar{f}_n(t,\xi)\to \bar{\psi}(t,\xi)
\]
in $L^p(\Omega)$ as $N\to \infty$, which further leads to the moment representation \eqref{eq:mmre}.~$\Box$

For fixed $m_1,\ldots,n_N\in \mathbb{N}$, assuming that
\[
\sum_{i=1}^M m_i+\sum_{j=1}^n n_j=2k,
\]
for some $k\in\mathbb{N}$,  we write 
\begin{eqnarray}\label{eq:limMM}
&&\E\{ \bar{f}_{m_1}\ldots \bar{f}_{m_M}\bar{f}_{n_1}^*\ldots\bar{f}_{n_N}^*\}=
\frac{1}{i^{\sum_{i=1}^M m_i}}\frac{1}{(-i)^{\sum_{j=1}^N n_j}}\frac{1}{\prod_{i=1}^M m_i!\prod_{j=1}^N n_j!}\\
&&~~~~~~~~\times \int_{\R^{(M+N)d}}dxdy\prod_{i=1}^M\phi_0(x_i)e^{-i\xi\cdot x_i}\prod_{j=1}^N \phi_0^*(y_j)e^{i\xi\cdot y_j}\E\{\V_t(x_1)^{m_1}\ldots \V_t(y_N)^{n_N}\}.
\nonumber
\end{eqnarray}
Since 
\[
\V_t(x)^m=\int_{[0,t]^m}\prod_{i=1}^m \dot{W}(s_i,x)ds=m!\int_{\Delta_m(t)}\prod_{i=1}^m \dot{W}(s_i,x)ds,
\]
we have
\begin{eqnarray}\label{eq:limMM1}
&&\E\{ \bar{f}_{m_1}\ldots \bar{f}_{m_M}\bar{f}_{n_1}^*\ldots\bar{f}_{n_N}^*\}=\frac{1}{i^{\sum_{i=1}^M m_i}}
\frac{1}{(-i)^{\sum_{j=1}^N n_j}}\int_{\R^{(M+N)d}}dxdy\prod_{i=1}^M\phi_0(x_i)e^{-i\xi\cdot x_i}
\prod_{j=1}^N \phi_0^*(y_j)e^{i\xi\cdot y_j}\nonumber\\
&&~~~~~~~~~~~~
\times \E\Big\{\int_{\Delta_{m,n}(t)}\prod_{i=1}^M
\prod_{k_1=1}^{m_i}\dot{W}(s_{i,k_1},x_i)ds_{i,k_1}\prod_{j=1}^N\prod_{k_2=1}^{n_j}\dot{W}(u_{j,k_2},y_j)du_{j,k_2}\Big\}.
\end{eqnarray}
Using the rules of computing the $2k-$th joint moment of mean-zero Gaussian random variables, we obtain
\begin{equation*}
\begin{aligned}
&\E\{\prod_{i=1}^M\prod_{k_1=1}^{m_i}\dot{W}(s_{i,k_1},x_i)\prod_{j=1}^N\prod_{k_2=1}^{n_j}\dot{W}(u_{j,k_2},y_j)\}\\
=&\sum_\F \prod_{(v_l,v_r)\in\F} \RR(v_l-v_r,z_l-z_r)=\sum_\F\int_{\R^{kd}}\prod_{(v_l,v_r)\in\F} 
e^{-\mu|w_l|^{2\beta}|v_l-v_r|}\frac{a(0)}{|w_l|^{2\gamma+d-2}}e^{iw_l\cdot(z_l-z_r)}\farc{dw_l}{(2\pi)^{d}},
\end{aligned}
\end{equation*}
where the summation  extends over all pairings $\F$ formed over the vertices
\[
\{s_{1,1},\ldots,s_{1,m_1},\ldots,s_{M,1},\ldots,s_{M,m_M}u_{1,1},
\ldots,u_{1,n_1},\ldots,u_{N,1},\ldots,u_{N,n_N}\}.
\]
Here, $v_l,v_r$ are the two vertices of a given pair, and $z_l,z_r$ are the corresponding $x,y$ variables. 
Now,~\eqref{eq:limMM1} becomes
\begin{eqnarray}
&&\E\{ \bar{f}_{m_1}\ldots \bar{f}_{m_M}\bar{f}_{n_1}^*\ldots\bar{f}_{n_N}^*\}
=\frac{1}{i^{\sum_{i=1}^M m_i}}\frac{1}{(-i)^{\sum_{j=1}^N n_j}} \sum_\F
\int_{\Delta_{m,n}(t)}dsdu\int_{\R^{(M+N+k)d}}dxdy\nonumber\\
&&\times\ \prod_{i=1}^M\phi_0(x_i)e^{-i\xi\cdot x_i}\prod_{j=1}^N \phi_0^*(y_j)e^{i\xi\cdot y_j}
\prod_{(v_l,v_r)\in\F} e^{-\mu|w_l|^{2\beta}|v_l-v_r|}\frac{a(0)}{|w_l|^{2\gamma+d-2}}e^{iw_l\cdot(z_l-z_r)}
\frac{dw_l}{(2\pi)^{d}}.
\label{eq:limMM2}  
\end{eqnarray}
Since each $w_l$ corresponds to a given pair $(v_l,v_r)$, we introduce $w_r$ and write 
\[
e^{iw_l\cdot(z_l-z_r)}=e^{iw_lz_l}e^{iw_rz_r}
\]
under the constraint 
\begin{equation}\label{aug306}
w_l+w_r=0.
\end{equation}
We say that $w_l$ corresponds to $v_l$ and $w_r$ corresponds to $v_r$. If a $w$ variable corresponds to 
some~$s_{i,j}$, we denote it as $p_{i,j}$; if it corresponds to some $u_{i,j}$, we denote it as $-q_{i,j}$. Therefore, we can write 
\[
\prod_{(v_l,v_r)\in\F}e^{iw_lz_l}e^{iw_rz_r}=\prod_{i=1}^M e^{ix_i\cdot(p_{i,1}+\ldots+p_{i,m_i})}\prod_{j=1}^N e^{-iy_j\cdot (q_{j,1}+\ldots+q_{j,n_j})},
\]
and an integration in $x,y$ in \eqref{eq:limMM2} leads to
\begin{equation}
\begin{aligned}
&\E\{ \bar{f}_{m_1}\ldots \bar{f}_{m_M}\bar{f}_{n_1}^*\ldots\bar{f}_{n_N}^*\}\\
=&\frac{1}{i^{\sum_{i=1}^M m_i}}\frac{1}{(-i)^{\sum_{j=1}^N n_j}}\sum_\F\int_{\Delta_{m,n}(t)}dsdu\int_{\R^{kd}}\prod_{(v_l,v_r)\in\F} e^{-\mu|w_l|^{2\beta}|v_l-v_r|}\frac{a(0)}{|w_l|^{2\gamma+d-2}}\frac{dw_l}{(2\pi)^{d}}\\
\times &\prod_{i=1}^M\hat{\phi}_0(\xi-p_{i,1}-\ldots-p_{i,m_i})\prod_{j=1}^N \hat{\phi}_0^*(\xi-q_{j,1}-\ldots-q_{j,n_j})\\
\end{aligned}
\label{eq:limMM3}
\end{equation}
under the constraint (\ref{aug306}) for all $(v_l,v_r)$. Comparing to \eqref{eq:limitJpc}, we see that the proof is complete.


\end{document}